\documentclass{PoS}
\usepackage{multirow}

\newcommand{\azz}{a^{0}_{0}(980)}
\newcommand{\fz}{f_{0}(980)}

\newcommand{\BR}{\mathcal{B}}

\newcommand{\ee}{e^+e^-}

\title{Isospin violations at BESIII}

\ShortTitle{Isospin violations at BESIII}

\author{\speaker{Hongrong Qi}\thanks{qihongrong@buaa.edu.cn} and Wencheng Yan\\
        for the BESIII Collaboration\\
        Beihang University, Beijing, 100083, People's Republic China\\
        E-mail: \email{qihongrong@buaa.edu.cn},
        \email{yanwc@buaa.edu.cn} }

\abstract{At present very large data samples in the energy region of 2.0-4.6 GeV were accumulated by the BESIII detector, which is operated in the upgraded Beijing electron positron collider (BEPCII). These data samples provide an unprecedented opportunity in the study of light hadron spectra and charmonium(-like) decays. We review some experimental analyses related to isospin violations at BESIII in this proceeding, which can be classify into three categories: isospin violating processes with a $f_0(980)$ or  $a^0_0(980)$ production, isospin violating processes with baryon final states, and isospin violating hadronic transitions in the charmonium system. }

\FullConference{XIII Quark Confinement and the Hadron Spectrum - Confinement2018\\
		31 July - 6 August 2018\\
		Maynooth University, Ireland}

\begin{document}

\section{Introduction}

Charmonium spectroscopy provides an ideal place to gain insights into the quark confinement and the decays of charmonium(-like) states. The isospin symmetry is conservative in strong interactions, while it could be violated in electromagnetic or weak interactions. The isospin violations will help us provide important probe of the relative amplitudes between strong interactions and electromagnetic interactions in charmonium decays, as well as the mass difference of the light $u$ and $d$ quarks.

The scalar meson $f_0(980)$, which is explained as a tetraquark~\cite{qq,qqqq}, a $K\bar{K}$ molecule~\cite{KK}, or a quark-antiquark gluon hybrid~\cite{gluon} in theory, is found with a width of $\sim10$ MeV in isospin violating processes~\cite{BESIII:2012aa,Ablikim:2015cob,a0f0mixing}, which is much narrower than that of $\sim50$ MeV in the isospin conservation decays~\cite{PDG2014}. The significant excess of $J/\psi \to \gamma \eta(1405) \to \gamma\pi^0 f_0(980)$ with the anomalously large isospin violation was observed at BESIII~\cite{BESIII:2012aa}. A triangle singularity mechanism is proposed to interpret this unusual phenomenon~\cite{Wu:2011yx}. Another theory, which is thought to be a feasible approach to clarify the phenomenon of the narrower $f_0(980)$, is the mixing mechanism of $a_0^0(980)-f_0(980)$~\cite{Achasov1979,Wu:2008hx}.
The BESIII collaboration recently observed the $a_0^0(980)-f_0(980)$ mixing with the statistical significance in excess of 5.0 standard deviations ($\sigma$)~\cite{a0f0mixing}.

Very recently, BESIII reported an evident difference in line shape and magnitude of the measured cross sections between $e^+e^-\to\Lambda(1520)(\to pK^-)\bar{n}K^{0}_{S}+c.c.$ and $e^+e^-\to pK^-\bar{\Lambda}(1520)(\to\bar{n}K^{0}_{S})+c.c.$~\cite{pKsnK}. We consider that such an isospin violating effect may be due to the interference between $I=1$ and $I=0$ final states. Other isospin violating processes with baryon final states, such as $J/\psi \to \Lambda\bar\Lambda \pi^0$ and $J/\psi \to \Lambda\bar\Sigma^0+c.c.$ , are also reported.

The isospin violating hadronic transitions in the charmonium system play an essential probe of the isospin asymmetry mechanism. For this case, we investigated three processes at BESIII: $\psi^\prime \to \pi^0 J/\psi$, $\psi^\prime \to \pi^0 h_c$, and $\chi_{c0,2} \to \pi^0 \eta_c$. In this proceeding, we also present the recently obtained results with isospin violations in so called exotic hadron decays at BESIII , such as $Y(4260) \to \pi^0 \eta J/\psi$ and $D^*_{s0}(2317)^\pm \to \pi^0 D^\pm_s$.
In addition, the doubly OZI suppressed decay $J/\psi \to \pi^0 \phi$ are also visited.

\section{Isospin violations with a $f_0(980)$ or $a_0^0(980)$ production}


The nature of the scalar states $\fz$ and $\azz$ have been controversial for several decades.
They are more inclined to be theoretically interpreted as tetra-quarks~\cite{qq,qqqq}, $K\bar{K}$ molecules~\cite{KK}, quark-antiquark gluon hybrids~\cite{gluon}, $etc.$, rather than conventional quark-antiquark mesons. Recent studies indicate that the isospin violating processes associated with the $f_0(980)$ or $a^0_0(980)$ would provide insights into the intrinsic nature of the unconventional mesons $f_0(980)$ and $a^0_0(980)$.

In 2012, a extremely narrow $f_0(980)$ signal with the width about 10 MeV via $J/\psi \to \gamma\pi^0 f_0(980)(\to\pi^{+,0}\pi^{-,0})$~\cite{BESIII:2012aa} was first observed, whereas its world average width is about $50-100$~MeV$/c^2$~\cite{PDG2012} in that year. In that analysis, it is found that the $f_0(980)$ predominantly come from the $\eta(1405)$ decay, $i.e.$, $\eta(1405)\to\pi^0 f_0(980)$~\cite{BESIII:2012aa}. According to the measurement, the ratio $R=(\eta(1405)\to f_0(980)\pi^0) \to \pi^+\pi^-\pi^0) / (\eta(1405)\to \azz \pi^0 \to \eta^0\pi^0\pi^0)$ is calculated to be $(17.9\pm4.2\footnote{In the proceeding, the single uncertainty is the quadratic sum of the statistical item and the systematic one; if there are two uncertainties, the first one is statistical, and second systematic.})\%$~\cite{BESIII:2012aa}, which is a considerably large isospin violation. In theory, a model of triangle singularity Feynman diagram was proposed to reconcile the anomaly~\cite{Wu:2011yx}.

In 2015, another $J/\psi \to V \pi^0 f_0(980)$ measurement was observed at BESIII~\cite{Ablikim:2015cob}, where the vector particle $V$ is the $\phi$ meson. Similarly, the observed $f_0(980)$ is narrower (about 15 MeV in width) than the world average in the same year~\cite{PDG2014}. It is strange that the significant $f_1(1285)$ instead of the $\eta(1405)$ is observed in the $\pi^0\fz$ invariant mass spectrum, and the $f_1(1285)\to\pi^0 f_0(980)$ is more clear with the statistical significance of 5.2~$\sigma$ in the $\fz\to\pi^+\pi^-$ channel~\cite{BESIII:2012aa}.


The mixing phenomenon in the $\fz-\azz$ system was first proposed in the late 1970s and expected to shed light on the nature of these two unconventional mesons~\cite{Achasov1979}. Taking into account the difference between the $K^+K^-$ and $K^0\bar{K}^0$ mass thresholds, the mixing mechanism predicted that a ``resonant" peak (actually a cusp effect) appears with a width of about 10-20 MeV in the isospin violating reaction $\pi^+\pi^-\to\pi^0\eta$~\cite{Achasov1979}, while the widths of both $\fz$ and $\azz$ are $50-100$~MeV$/c^2$~\cite{PDG2014,PDG} excluding in isospin violating decays. Later, theorists proposed to examine the  $\fz-\azz$ mixing mechanism in the isospin violating decays $J/\psi\to\phi\eta\pi^0$~\cite{Wu:2007jh}
and $\psi'\to\gamma\chi_{c1}\to\gamma\pi^+\pi^-\pi^0$~\cite{Wu:2008hx}. In addition, mixing intensities, $\xi_{fa}$ and $\xi_{af}$, defined as effective experimental probes for the nature of $\azz$ and $\fz$, are sensitive to the coupling constants for $\azz\to K\bar{K}$ and $\fz\to K\bar{K}$, respectively.

In 2018, BESIII reported the observation of the $\fz-\azz$ mixing phenomenon in the decays of $J/\psi\to\phi\eta\pi^0$ ($\eta\to\gamma\gamma$ and $\eta\to\pi^+\pi^-\pi^0$, $\phi\to K^+ K^-$) and $\psi'\to\gamma\chi_{c1}\to\gamma\pi^+\pi^-\pi^0$~\cite{a0f0mixing}. The excesses of $\fz$-$\azz$ mixing are determined with a statistical significance of larger than 5$\sigma$ for the first time, and the corresponding absolute branching fractions and mixing intensities are summarized in Table~\ref{a0f0 sum}.

 \begin{table*}[!htbp]
  {\caption{{\small{ The branching fractions ($\BR$) and the intensities ($\xi$) of the $\fz$-$\azz$ mixing. The first uncertainties are statistical, the second ones are systematic, and the third are related to parameterization of $\fz$ and $\azz$.}}}
  \label{a0f0 sum}}
  \begin{footnotesize}
  \begin{tabular}{lccccccccc}
  \hline \hline
  \multirow{2}{*}{Channel}      &                   \multicolumn{2}{c}{$\fz\to\azz$}               &      \multirow{2}{*}{$\azz\to\fz$}    \\
                            &~~~~~~~~{Solution I}~~~~~~~~~&~~~~~~~~{Solution II}~~~~~~~ &                                   \\ \hline
  $\BR$(mixing)~$(10^{-6})$   &~~~$3.18\pm0.51\pm0.38\pm0.28$~~~ & ~~~$1.31\pm0.41\pm0.39\pm0.43$~~~ &~~~$0.35\pm0.06\pm0.03\pm0.06$~~~      \\
  $\BR$(EM)~$(10^{-6})$      & $3.25\pm1.08\pm1.08\pm1.12$ & $2.62\pm1.02\pm1.13\pm0.48$ &                ---                \\
  $\BR$(total)~$(10^{-6})$   & $4.93\pm1.01\pm0.96\pm1.09$ & $4.37\pm0.97\pm0.94\pm0.06$ &                ---                \\
  $\xi$ (\%)                & $0.99\pm0.16\pm0.30\pm0.09$ & $0.41\pm0.13\pm0.17\pm0.13$ &  $0.40\pm0.07\pm0.14\pm0.07$      \\
  \hline \hline
  \end{tabular}
  \end{footnotesize}
 \end{table*}

\begin{figure}[htbp]
 \centering
 \vskip -0.3cm
 \includegraphics[width=0.70\textwidth]{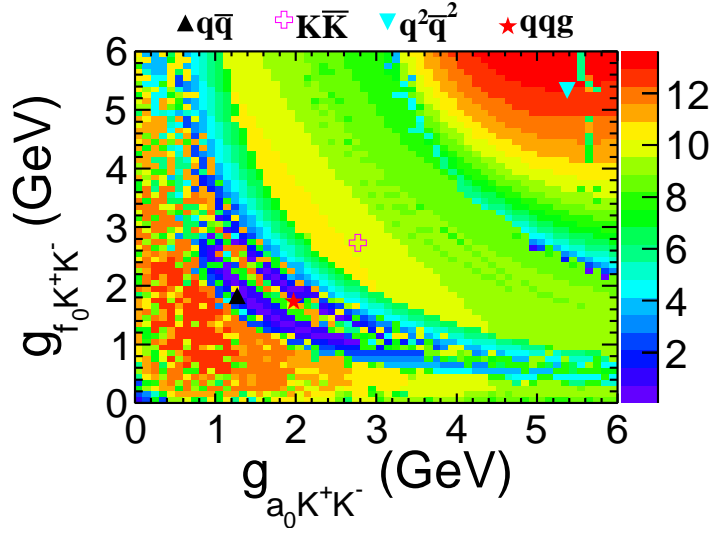}
 \vskip -0.2cm
 \caption{The statistical significance of the signal scanned in the two-dimensional space of $g_{a_{0}K^{+}K^{-}}$ and $g_{f_{0}K^{+}K^{-}}$. The regions with higher statistical significance indicate larger probability for the emergence of the two coupling constants. The markers indicate predictions from various illustrative theoretical models. }
\label{a0f0_fig}
\end{figure}

We obtained the statistical significance of the mixing signal by scanning the two coupling constants $g_{a_{0}K^{+}K^{-}}$ and $g_{f_{0}K^{+}K^{-}}$, which is depicted in Fig.~\ref{a0f0_fig}. For this two-dimensional distribution, the regions with higher statistical significance indicate larger probability for the emergence of the two coupling constants. The predicted coupling constants from various models~\cite{Wu:2008hx} are displayed as well (color markers), but the theoretical uncertainties on the models are not considered here.
Due to the limited statistics, a solid conclusion on internal structure of these two unconventional mesons is not yet drawn.

\section{Isospin violations in baryon final states}

\begin{figure}
  \centering
  \includegraphics[scale=0.5]{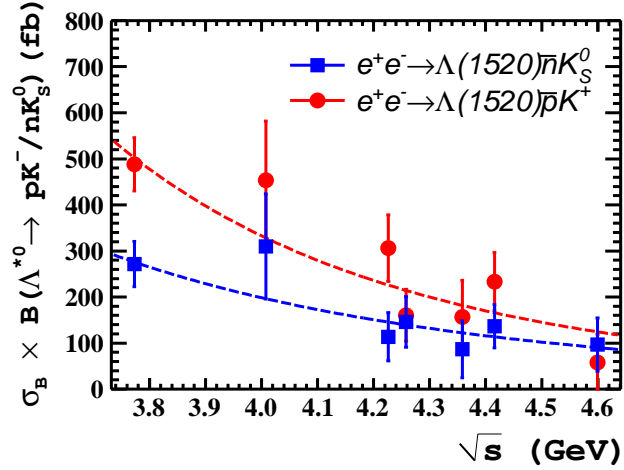}\\
  \caption{(Color online)
    Distributions of $\sigma_{\rm B}[e^+e^- \to \Lambda(1520) \bar{n}K^{0}_{S}]\times \BR[\Lambda(1520) \to p K^-]$,
    and $\sigma_{\rm B}[e^+e^- \to \Lambda(1520) \bar{p}K^{+}]\times\BR[\Lambda(1520) \to n K_S^0]$
    versus c.m. energy.
    The dots with error bars, which are the combined statistical and uncorrelated systematic uncertainties, represent data.
    The blue and red dashed lines are the fits to the cross sections
    of $e^+e^- \to \Lambda(1520) \bar{n}K^{0}_{S}$ and $e^+e^- \to
    \Lambda(1520) \bar{p}K^{+}$, respectively.}\label{fig:cross-section-cor}
\end{figure}

Recently, BESIII reported that there is an evident difference in the line shape and magnitude of the measured Born cross sections between $\ee\to\Lambda(1520)(\to pK^-)\bar{n}K^{0}_{S}${\footnote{In this section, charged conjugated modes are included unless otherwise indicated.}}
and $\ee\to pK^-\bar{\Lambda}(1520)(\to\bar{n}K^{0}_{S})$, as shown in Fig.~\ref{fig:cross-section-cor}~\cite{pKsnK}. The statistical significance of the cross-section difference is 3.1$\sigma$ at the c.m. energy of 3.770 GeV, while less than 3$\sigma$ at other energies due to lower statistics.
Such an isospin violation effect is probably due to the interference between $I=1$ and $I=0$ final states.
The final states $\Lambda(1520)\bar{n}K^{0}_{S}$ and $\Lambda(1520)\bar{p}K^+$ can
be produced from $\bar{p}K^+$ and $\bar{n}K^0_S$ systems either in $I=1$ or
$I=0$ states, namely, excited $\bar{\Sigma}^*$ or $\bar{\Lambda}^*$ states.
These two states can decay into both $\bar{p}K^+$ and $\bar{n}K^0_S$ final states,
but with a sign difference from $CG$ coefficients.
Another possible approach is $\ee \to \bar{K}^*K$ with the highly excited $\bar{K}^*$ decaying into $\Lambda(1520)(\to pK^-)\bar{n}$
or $\Lambda(1520)(\to nK^0_S)\bar{p}$.  The $\bar{K}^*K$ system can be produced from $I=1$
(excited $\rho^*$) or $I=0$ (excited $\omega^*$ or $\phi^*$) states,
where the interference effect can occur.  If the final state is
$p\bar{N}^*$ or $n\bar{N}^*$, a similar pattern could be observed.
More experimental data are desirable to confirm these interpretations and
speculations in the future.
Theoretically, Ref.~\cite{Cao:2018kos} claimed that the formalism associated with the electromagnetic form factors of the $\Lambda(1520)$ will shed light on the isospin violating cross sections if the measurement will be improved with higher precision.

In this proceeding, we present the recent measurements at BESIII of the branching fractions~\cite{Ablikim:2012uqz} of the isospin violating and isospin conserving decays of $J/\psi(\psi')\to\Lambda\bar\Lambda\pi^0$ and $J/\psi(\psi')\to\Lambda\bar\Lambda\eta$, with drastically more precision than previous results~\cite{PDG2012}.
The branching fractions in $J/\psi$ decays are determined to be
$\BR(J/\psi\to\Lambda\bar\Lambda\pi^0)=(3.78\pm0.27\pm0.30)\times10^{-5}$ and
$\BR(J/\psi\to\Lambda\bar\Lambda\eta)=(15.7\pm0.80\pm1.54)\times10^{-5}$.
Therefore, the ratio of $\BR(J/\psi\to\Lambda\bar\Lambda\pi^0)$ to $\BR(J/\psi\to\Lambda\bar\Lambda\eta)$
is equivalent to 0.24, which indicates extremely large isospin violations.
The branching fraction of $\psi'\to\Lambda\bar\Lambda\eta$ is measured to be $(2.48\pm0.34 \pm0.19)\times10^{-5}$, while the upper limit of the branching fraction $\BR(\psi'\to\Lambda\bar\Lambda\pi^0)<0.29\times10^{-5}$ is obtained at the 90\% confidence level.
The ratio $R=\BR(\psi'\to\Lambda\bar\Lambda\pi^0) / \psi'\to\Lambda\bar\Lambda\eta<0.12$.

The branching fraction of the isospin symmetry breaking decay $J/\psi\to\Lambda\bar\Sigma^0$ will help
elucidate the decay amplitude of $J/\psi\to B_8 \bar{B}_8$ via three gluons or one photon~\cite{Kowalski:1976mc,Kopke:1988cs}, and provide an insight into the SU(3)-flavour asymmetry~\cite{Kopke:1988cs,Zhu:2015bha} as well,
where $B_8$ ($\bar{B}_8$) denotes an octet baryon (antibaryon).
Using the sample of $(225.2\pm2.8)\times10^6$ $J/\psi$ events collected at BESIII,
the branching fractions of $J/\psi\to\Lambda\bar\Sigma^0$ and $J/\psi\to\bar\Lambda\Sigma^0$ are measured to be $(1.46\pm0.11\pm0.12)\times10^{-5}$
and $(1.37\pm0.12\pm0.11)\times10^{-5}$, respectively~\cite{Ablikim:2012bw}.

\section{Isospin violating hadronic transitions in the charmonium system}

The hadronic transitions $\psi^\prime \to \pi^0(\eta) J/\psi$ were suggested
to be a reliable source for the extraction of the light quark mass ratio $m_u/m_d$, obtained by the equation~\cite{Ioffe:1979rv,Ioffe:1980mx},
$$R_{\pi^0/\eta}=3\left(\frac{m_d-m_u}{m_d+m_u}\right) \frac{F^2_\pi}{F^2_\eta} \frac{M^2_\pi}{M^2_\eta} \left|\frac{\overrightarrow{q_\pi}}{\overrightarrow{q_\eta}}\right|^3,$$
where $R_{\pi^0/\eta}$ is the ratio of $\BR(\psi^\prime \to \pi^0 J/\psi)$ to $\BR(\psi^\prime  \to \eta J/\psi)$,
$F_{\pi(\eta)}$ and $M_{\pi(\eta)}$ are the decay constant and mass of
the $\pi(\eta)$, respectively,
$\overrightarrow{q}_{\pi(\eta)}$ stands for the momentum of the $\pi(\eta)$ in the rest
frame of the $\psi^\prime$.
Because the $J/\psi$ and $\psi^\prime$ are SU(3) singlets, it is obvious
that the decay $\psi^\prime \to \pi^0 J/\psi$ violates isospin symmetry,
and the decay $\psi^\prime  \to \eta J/\psi$ violates SU(3) flavor symmetry~\cite{eta-note, Guo:2009wr}.
Thereby, the decay amplitudes reflect the flavor symmetry breaking, and it is the mass differences within the multiplet that generates the isospin or the SU(3) breaking~\cite{Guo:2009wr}.
The ratio $R_{\pi^0/\eta}$ was found to be $(3.74\pm0.06 \pm0.04)\%$ by BESIII, with unprecedented high precision~\cite{Ablikim:2012aq}.

Another isospin-violating transition $\psi^\prime \to \pi^0 h_c$ was observed to be $\BR(\psi^\prime \to \pi^0 h_c) = (8.4\pm1.3\pm1.0) \times 10^{-4}$ by the BESIII Collaboration~\cite{Ablikim:2010rc}.
The measurements is consistent with a simple single-channel calculations based on the QCD multipole expansion using the Cornell potential model with predicting a partial width of (0.12-0.36) KeV~\cite{Kuang:1988bz}.

BESIII Collaboration also search for the isospin-violating $\pi^0$ transition $\chi_{c0,2}\to\pi^0\eta_c$, but only the upper limits on the branching fractions are determined at the 90\% confidence level to be $\BR(\chi_{c0}\to\pi^0\eta_c)<1.6\times10^{-3}$ and $\BR(\chi_{c2}\to\pi^0\eta_c)<3.2\times10^{-3}$~\cite{Ablikim:2015rla}.

\section{Other isospin violations at BESIII}

In recent years, BESIII reported a number of $XYZ$ particles (also known as the charmonium-like states) in the charmonium region.
Some isospin violating processes associated with unconventional hadrons are searched for or observed, such as $Y(4260) \to \pi^0 \eta J/\psi$, $D^*_{s0}(2317)^\pm \to \pi^0 D^\pm_s$.
The tetraquark model~\cite{Faccini:2013lda} predicts that $Z_c(39000)$ can be produced in $Y(4260)\to\pi^0 \eta J/\psi$ with $Z_c(3900)$ decaying into $\pi^0 J/\psi$ and possibly $\eta J/\psi$ in the presence of sizable isospin violation.
The molecular model~\cite{Wu:2013onz} predicts a peak in the cross section of $Y(4260)\to\pi^0 \eta J/\psi$ at the $D\bar{D}_1$ threshold, and a narrow peak in the $\eta J/\psi$ invariant mass spectrum at the $D\bar{D}^*$ threshold.
BESIII performed the isospin violating process $Y(4260) \to \pi^0 \eta J/\psi$, but no signal excess in either the $\eta J/\psi$ mass spectrum or the cross-section line shape~\cite{Ablikim:2015xfo}.
The upper limit on the ratio of the branching fractions ($Y(4260) \to \pi^0 \eta J/\psi$)/($Y(4260) \to \pi^0 \pi^0 J/\psi$) at the 90\% confidence level is 0.15 at $\sqrt{s}=4.226$ GeV and 0.65 at $\sqrt{s}=4.257$ GeV~\cite{Ablikim:2015xfo}.

The $D^*_{s0}(2317)^+$ meson is generally thought to be the P-wave $c\bar{s}$ state with spin-parity $J^P=0^+$.
With a mass that is 45 MeV$/c^2$ below the $DK$ mass threshold, it is proposed as a $DK$ molecule~\cite{vanBeveren:2003kd}, a $c\bar{s}q\bar{q}$ tetraquark state~\cite{Cheng:2003kg}, one of the chiral charmed doublets~\cite{Nowak:2003ra}, or a mixture of a $c\bar{s}$ meson and a $c\bar{s}q\bar{q}$ tetraquark~\cite{Terasaki:2003qa}.
BESIII observed the $D^*_{s0}(2317)^\pm$ in the decays $D^*_{s0}(2317)^\pm \to \pi^0 D^\pm_s$, and the absolute branching fraction is measured to be $\BR(D^*_{s0}(2317)^\pm \to \pi^0 D^\pm_s)$
=$1.00^{+0.00+0.00}_{-0.14-0.14}$ for the first time~\cite{Ablikim:2017rrr}.

In addition, the doubly OZI suppressed decay $J/\psi \to \pi^0 \phi$ are also investigated at BESIII.
Considering the interference between $J/\psi \to \pi^0 \phi$ and $J/\psi \to \pi^0 K^+ K^-$,
the branching fraction  $J/\psi \to \pi^0 \phi$ is determined as $(2.94\pm0.16\pm0.16)\times10^{-6}$ for constructive interference and $(1.24\pm0.33\pm0.30)\times10^{-7}$ for destructive interference~\cite{Ablikim:2015mua}.

\section{Summary}
In generally, there are two possible sources of isospin symmetry breaking, namely
the electromagnetic processes and the difference between the masses of up and down
quarks. Of late years, in order to interpret the phenomena with large isospin violations, a mixing mechanism and a triangle singularity diagram were proposed.
In these proceeding, we summarized the isospin violating measurements in recent BESIII analyses with respect to these theories, which are mainly divided into three categories: isospin violating processes with a $f_0(980)$ or $a_0^0(980)$ production, isospin violations with baryon final states, and isospin violating hadronic transitions in the charmonium system.



\begin{thebibliography}{99}
\bibitem{eta-note}
Here we assume the $\eta$ is in the SU(3) octet, and the effect
of the $\eta-\eta^\prime$ mixing is assumed to be small~\cite{Casalbuoni:1992fd}.
\end{thebibliography}
\end{document}